\documentclass[preprint,fleqn,10pt,prb,balancelastpage,twocolumn]{revtex4}
\usepackage{amsmath}
\usepackage{graphicx}

\input{tcilatex}

\begin{document}

\title{Magnetic short range order above the Curie temperature in Fe and Ni}
\author{Vladimir Antropov}
\affiliation{Condensed Matter Physics Dept., Ames Lab, Ames, IA, 50011}

\begin{abstract}
We report the first results of time-dependent density functional simulations
of magnetic properties of Fe and Ni at finite temperatures. They reveal the
existence of local moments in Ni above the Curie temperature, coupled to
strong short range order. In Fe the short range order is also present to a
lesser extent but spin wave like excitations persist in both materials well
above the Curie temperature in the paramagnetic state. The prevailing view
of magnetic order should be reconsidered in light of these findings.
\end{abstract}

\maketitle

A proper description of the microscopic behavior of magnetic systems at
finite temperature represents one of a few fundamental problems still
existing in the theory of magnetism. While numerous model theories with
different adjustable parameters lead to some description of magnets at
finite temperatures, the attendant approximations significantly influence
the final results. \emph{Ab initio} methods have been used since the early
1980's; however they are limited to zero temperature and largely reproduce
the Heisenberg model. As might be expected, such methods provide a good
description of magnets with localized moments, but the results for itinerant
magnets are rather unsatisfactory. In particular, the temperature dependence
of longitudinal fluctuations and magnetic short-range order (MSRO) have
always been neglected.

MSRO in itinerant magnets is the subject of a long-standing controversy.
Early reports on inelastic neutron scattering experiments indicated the
persistence of spin wave-like modes with relatively small wave vectors above
the Curie temperature $T_{c}$ in both Ni and Fe\cite{MSRO}. Based on these
results, the local band theory was developed\cite{Korenman}, but its
assumptions are incompatible with the observed Curie-Weiss susceptibility
and specific heat anomaly\cite{EDWARDS}. Since then, various experiments\cite%
{SRO} confirm the presence of some degree of MSRO in many ferromagnets, and
the present belief is that it can be very strong. However, up to now theory
has failed to estimate the degree of MSRO in different magnets, owing to the
formidable challenges in implementing a tractable theory for real materials.

Several quantitative theoretical methods based on the use of the calculated
band structure have been developed. In many cases the \emph{ab initio}
version of the single-site coherent potential approximation (CPA) was
employed. The first calculation\cite{Oguchi} evaluates the MSRO in Fe by
calculating the non-local exchange coupling. However, the single-site CPA
presupposes that MSRO is absent; hence the calculated MSRO is small. Further
CPA calculations\cite{Onsager} also obtained a good description of
finite-temperature properties of bcc Fe, but do not find any local magnetic
moment (LMM) at high temperatures in fcc Ni. Using Onsager's reaction-field
approach\cite{Onsager}, the description of Ni can be improved, which
indicates that some interatomic correlations beyond the mean-field
approximation (MFA) should be included, at least for Ni.

Different tight-binding methods not based on the CPA have been used to
estimate the MSRO and the effective coupling in Fe and Ni \cite{TB}. Most
did not find strong MSRO and failed to explain the corresponding
experimental data. In Ref.~\cite{HEINELIXT} it was claimed that the
effective coupling is much stronger at the Curie temperature than at $T=0$,
which makes the temperature-dependence of the exchange coupling a very
important problem. However, as we will discuss below, the tight-binding
approximation leads to a strong underestimation of MSRO, and hence to
different physics. The first band-structure calculations of the effective
exchange coupling\cite{Korenman} may be regarded as being in the limit which
is opposite to the CPA, since the local band theory assumes strong MSRO.
However, these calculations\cite{WANG} yielded a much smaller degree of MSRO
in contradiction with the initial assumption. This contradiction has yet to
be resolved. Thus, it seems that no numerical calculations so far have
explicitly shown the presence of strong MSRO in Fe or Ni based on their band
structures, while the experiments clearly indicate its presence.

The second key point concerns the proper inclusion of longitudinal
fluctuations in itinerant magnets at finite-temperature, which are missing
in any RPA theory.

If it were not for serious challenges limiting its tractability, \emph{ab
initio} spin dynamics would be a very powerful tool to address all these
issues. In Ref.\cite{SD} a combination of first principles spin dynamics
with different stochastic or deterministic types of finite-temperature
description was proposed. In the present paper we present an implementation
that makes this technique practicable, and apply it for the first time to
studies of finite-temperature behavior of ferromagnetic Fe and Ni. This
approach is based on time-dependent spin density functional theory \cite%
{TDDFT}. The main equation in this theory is 
\begin{equation}
i\frac{\partial \psi }{\partial t}=\left( -\nabla ^{2}+V(\mathbf{r})-\frac{1%
}{2}(\mathbf{B}_{xc}(\mathbf{r})+\mathbf{B}_{ext}(\mathbf{r}))\mathbf{\sigma 
}\right) \psi .  \label{e1}
\end{equation}%
where $V(\mathbf{r})$ is a non magnetic part of the total potential, $%
\mathbf{B}_{xc}(\mathbf{r})$ is the magnetic part of the exchange
correlation potential and $\mathbf{B}_{ext}(\mathbf{r})$ is the external
magnetic field.

Instead of considering the evolution of the non-local density matrix, one
can introduce the closed system of coupled equations for the three local one
electron quantities: the charge density $n_{\nu }\left( \mathbf{r,}t\right) $%
, the magnetization density $\mathbf{m}_{\nu }\left( \mathbf{r,}t\right) $,
and the velocity density $\mathbf{v}_{\nu }\left( \mathbf{r},t\right) $.
Separating the magnetization dynamics adiabatically, one can obtain the main
one-electron equation of spin dynamics 
\begin{equation}
\frac{d\mathbf{m}_{\nu }}{dt}=\gamma \mathbf{m}_{\nu }\times \left[ \mathbf{B%
}_{\nu }^{kin}(\mathbf{r,}t)+\mathbf{B}_{xc}(\mathbf{r,}t)+\mathbf{B}_{ext}(%
\mathbf{r,}t)\right]  \label{kss}
\end{equation}%
where $\mathbf{B}_{\nu }^{kin}(\mathbf{r})=\mathbf{\nabla }\left( n_{\nu }%
\mathbf{\nabla m}_{\nu }\right) /n_{\nu }$, $d/dt=\partial /\partial t+%
\mathbf{v}_{\nu }\mathbf{\nabla }$, and $\mathbf{v}_{\nu }$ is the velocity
field for the electron liquid.

The basic assumption of the local spin density approximation (LSDA) is that
the exchange-correlation potential of the homogeneous electron gas with
fixed charge density $n$ and magnetization density $\mathbf{m}$ can be
applied to real external potentials. In this approximation $\mathbf{B}_{xc}(%
\mathbf{r})\uparrow \uparrow $ $\mathbf{m}(\mathbf{r})$, and hence assumes
very strong exchange correlation field $\mathbf{B}_{xc}(\mathbf{r,}t)$
compared to the `kinetic' field $\mathbf{B}^{kin}(\mathbf{r,}t)$ and the
smallness of dynamics associated with the exchange correlation term. This
important approximation assumes that correlation effects do not contribute
to the spin dynamics of magnets, but attributes all of the dynamics to the
kinetic term. Many static calculations of spin excitations have generally
confirmed that this approximation is valid for many materials. In Ref. \cite%
{SAGA} we estimated the strength of correlation effects by calculating the
contribution of non-local Coulomb correlations to the spin wave stiffness in
the GW approximation. This contribution happened to be very small compared
to the kinetic contribution. Below we will confirm that this approximation
is quite reasonable for dynamic simulations as well.

In this paper we solve Eq.(\ref{kss}) by using the classical Langevin
approach for finite temperatures, and employ the so-called orbital dynamics
approach\cite{SD} so that each orbital can precess with its own frequency,
independent of the other orbitals. In this treatment we avoid the standard
adiabaticity which is assumed in the dynamic rigid spin approximation, and
account for all decoherence effects properly within LSDA. This approach is
similar to the linear response technique when the external field is small.
However, our approach does not necessarily assume that the \emph{perturbation%
} is small.

All calculations were performed within LSDA using the tight-binding linear
muffin-tin orbital method. In typical calculations with 100-120 atoms per
cell we used 20-30 $k$-points in the irreducible part of the Brillouin zone.
The correct description of possible long-range magnetic order (especially at
low temperatures) requires that the simulation cell contains many atoms. To
avoid this numerical difficulty, we added temperature-dependent spin-spiral
boundary conditions \cite{SS} that is dynamical, with its corresponding
equation of motion. Similar to the static case, such combined
real/reciprocal space spin dynamics allowed us to significantly reduce the
number of atoms per cell in our simulations and made numerous statistical
calculations possible. From the physical point of view, the simultaneous
inclusion of real-space short-range modes (inside the supercell) and
long-wavelength modes (between supercells) enables the description of
magnetic excitations developing at different length scales, as well as their
interactions, on an equal footing. As we now show, the ability to include
multiple-length scales was essential to this work.

First, let us discuss static properties. In Fig.1 we show the total
magnetization of Fe as a function of temperature. The results for 128-atom
simulations (using dynamical spin spiral boundary conditions) for Fe are
shown together with the finite-size scaling result. The finite-size scaling
results were obtained using supercells of four different sizes for both Ni
and Fe. The effectiveness of the dynamical spiral boundary conditions can be
seen in the 128-atom simulation. The low magnetization 'tail' above $T_{c}$
was reproduced in standard real space calculations (Monte Carlo simulations
for the Heisenberg model) only when 7000-8000 atoms were used\cite{Fisher}. $%
T_{c}$ resulting from the simulation is quite satisfactory. For Fe, nearly
perfect agreement with experiment is obtained: 1070K (1023K); in Ni $T_{c}$
is about 25\% smaller then experiment: 470 K (623K). However, this
discrepancy may be attributed to the LSDA itself: the total energy
difference between fully ferromagnetic and nonmagnetic Ni is about 550K
(with all parameters identical to the dynamical simulation). While this
upper limit to the $T_{c}$ in Ni is somewhat higher when gradient
corrections are added to LSDA, we do not at present have a sufficiently
accurate density functional to disentangle the LSDA contribution to the
error. We only suggest that this error is connected with some on-site
Coulomb correlations which are missing in LSDA at $T=0$ K. The proper
acccount of quantum corrections could significantly affect $T_{c}$ as well.

The effective paramagnetic moment $p_{c}$ the was extracted from the
calculations of high-temperature susceptibility. These moments compared to
experimental values\cite{Fe} are 3.3 (3.13) and 1.75 (1.6) Bohr magnetons
for Fe and Ni respectively.

To interpret these large moments we will now discuss the dynamical
properties. A very unusual picture of magnetic order near $T_{c}$ emerges.
The crucial physical quantity which reveals a degree of short- or long-range
order in magnets is the spin-spin correlation function%
\begin{equation}
S_{ij}\sim \left\langle \left\langle \mathbf{m}_{i}\left( t\right) \mathbf{m}%
_{j}\left( t^{\prime }\right) \right\rangle \right\rangle ,  \label{e4}
\end{equation}%
where $\left\langle \left\langle ...\right\rangle \right\rangle $ indicates
thermodynamic average. The short-range part of $S_{ij}$ is strongly affected
by the presence of itineracy in the magnetic system. The analysis of $S_{ij}$
shows that the average angle between nearest neighbors' LMM is 67$^{0}$ in
bcc Fe near $T_{c}$, however in fcc Ni it is just 28$^{0}$. These numbers
are very stable as a function of temperature: within a temperature range of
0.95$T_{c}$ to 1.05$T_{c}$ their overal changes do not exceed 3-4\%. The
deviation from 90$^{0}$ indicates that in Fe the MSRO is relatively small
(but noticeable), while in Ni it is enormous. A real-space analysis of the
correlation function shows that in Ni the dynamic ordering is of the
spin-spiral or domain-wall type, with a period of $\sim $14 a.u. On the
other hand, in Fe nearest neighbor correlations dominate and a relatively
weak superparamagnetic type MSRO is realized at $T_{c}$. Overall, the
dynamic correlation function has very well defined peaks well above $T_{c}$
and clearly demonstrates the persistence of the spin-wave-like perturbations
in the paramagnetic state of Fe and Ni. The relatively high energy of
short-wave ferromagnons in these materials is primarily responsible for this
feature.

The presence of such MSRO at $T_{c}$ indicates that the nearest neighbor
interactions in these materials must be larger then $T_{c}$. However, the
exchange couplings in these materials are well known (see Ref.\cite%
{WANG,PRB1}) and corresponding numbers are somewhat lower then $T_{c}$
making the existence of the strong MSRO\ in these materials impossible. On
the other hand, starting from pioneering calculations in Ref.\cite{WANG} all
calculations of $J_{ij}$ assumed a long wave approximation (LWA) which has
smallness parameter $\Delta \approx J_{ij}/I$ , where $I$ is the Stoner
parameter\cite{JMMMme}. In general, this approximation is inapplicable to
systems with strong interatomic interactions (as in the case of itinerant
magnets). We found that the usage of the more general definition $\left(
J=\chi ^{-1}\right) $ leads to a much better interpretation of our results.
First of all, the LWA is suitable for a relatively `localized' system such
as Fe, and the corresponding change in the nearest neighbor exchange is
relatively small ($\sim $15\%) but still somewhat higher than $T_{c}$. Ni
represents a rather itinerant system, and any local approach (especially
LWA) is expected to produce a large error. In our case, $J_{01}$ for Ni is
increased by more than 3 times relative to the LWA, clearly that the
effective coupling is much larger than $T_{c}$. These results explain why
the MSRO was missed in previous calculations of $J_{ij}$ and indicate that
the traditional MF approach (or any other approach which is based on the
assumption of 'no short-range order') is inapplicable to itinerant systems
in general, predicting very high $T_{c}$ (above 1000 K in Ni). This
interpretation has a qualitative character due to known violation of
adiabaticity criteria in the itinerant magnets.

The MSRO allows us to provide a different interpretation of the
susceptibility results as described above. The increase of the effective
magnetic moment $p_{c}$ in the Curie-Weiss law compared to the localized
model is directly related to clustering (for example, of the
superparamagnetic type) when the effective moment of the cluster is several
times larger then original atomic moment. The ratio of the effective moment
to the atomic moment is larger in Ni than in Fe. Our calculations also
indicate that in Ni a significant contribution (15\%) comes from
longitudinal fluctuations. The high temperature paramagnetic state of Ni has
both tangential and longitudinal modes as soft modes which are much closer
to each other than at $T=0$ K. We performed several model calculations for
Ni assuming static disorder with 90$^{0}$ average angle between nearest LMM
(no MSRO). For all eight random configurations studied the value of $p_{c}$
was much lower (about 50-60\% of our theoretical value), clearly supporting
MSRO contribution to the $p_{c}$. In addition, all final self consistent
states in Ni for these random configurations appeared to be non magnetic.
So, we believe that strong MSRO is a necessary requirement for fcc Ni to be
ferromagnetic and a similar conclusion should apply to all itinerant magnets.

Such strong MSRO can not appear only at finite temperatures; it must exist
already at $T=0$ K (as indicated by our calculations of the effective
coupling). To study MSRO in more detail we performed the following model
calculations at T=0 K. First, we studied the degree of localization by
rotating the LMM at one site by an arbitrary angle (similar to Ref.\cite{rot}
but with LWA removed). In case of the fully localized Heisenberg type of
interaction one can expect that such perturbation is perfectly stable and
the total energy is described by $\left( 1-cos\theta \right) $ type of
behavior. Our calculation revealed completely different picture (Fig.2). In
case of single site perturbation, the atomic magnetic moment in Fe
disappears at approximately 120$^{0}$. Longitudinal fluctuations at this
point become so important that the moment is inverted. As we mentioned
above, such effects can not be described in any tight-binding type of
approach. In Ni the situation is even more complicated and at $\sim $50$^{0}$
the local moment is completely destroyed. These calculations immediately
suggest that both Fe and Ni do not have a well defined LMM at finite
temperature. We also found, that the LMM with opposite spin configuration in
bcc Fe can be stabilized if the perturbation is extended to the first shell
of neighboring atoms, providing a clear range of minimal MSRO to support
such strong local perturbation. In the itinerant Ni the complete flip of LMM
can not be stabilized even if the perturbation is extended to the first
neigboring atoms (Fig.2). The perturbation is stabilized only when it is
extended to three shells of neighbors. These calculations strongly support
the idea of giant MSRO in fcc Ni and relatively small (but essential) MSRO
in bcc Fe.

Another indication of itineracy (and strong MSRO) only was obtained from
studies of non-magnetic configurations. We estimated the range of existence
of LMM in non magnetic fcc Ni. In case when LMM\ exist only at one atom such
fluctuation is unstable and LMM simply does not exist. However, if the LMM
is fixed at its equilibrium value (0.6 $\mu _{B}$) on nearest atoms, the
atomic LMM is stable. According to our LSDA calculations fcc Ni at its
experimental density has stable LMM of $\sim $0.6 $\mu _{B}$ only if it has
ferromagnetic interaction with two nearest shells of atoms with the same
moments. This result clearly indicates that magnetism in Ni has itinerant
character. Its nature in this material is not so much in intra-atomic
interactions but rather in interatomic ferromagnetic interactions with
nearest neighbors. This also indicates that the standard adiabatic criterion
is not applicable in Ni at finite temperatures.

Such strong MSRO also allow us to resolve a very old issue of the importance
of quantum corrections for $T_{c}$ and high-temperature susceptibility
calculations. These corrections are traditionally small in the localized
systems ($4f$ magnets), but always considered to be significant in the
itinerant systems. MSRO described above also strongly suppresses quantum
corrections in the case of the itinerant magnets by increasing the effective
moment.

In summary, we performed \textit{ab-initio} spin dynamic simulations at
finite temperatures using time-dependent spin density functional theory. Our
calculations revealed the existence of magnetic short range order at $T_{c}$
in $3d$ ferromagnetic magnets Fe and Ni. This short-range order is
relatively small in bcc Fe ( the average angle between local magnetic
moments $\left\vert \theta \right\vert \sim 70^{0}$), while in fcc Ni it is
rather large ( $\left\vert \theta \right\vert \sim 25^{0}$). The existence
of 'distorted' spin waves above $T_{c}$ significantly reduces the range of
applicability of the traditional models of magnetism which are based on the
idea of the spin wave disappearing at $T_{c}$, and demonstrates that more
reliable models of magnetism are needed. This MSRO strongly affects all
observable properties and establishes that the thermodynamics of Fe and Ni
can be described in the framework of the density functional approach. We
believe that MSRO is a natural property of any magnet over a very wide
temperature range and is the \textit{sine qua non} in low moment ($m<2\mu
_{B}$) itinerant magnets where MSRO dramatically influence thermodynamical
properties. This MSRO is unique in each system and has a certain 'memory'
about the ground state of that system. From the methodical point of view the
successful usage of our approach clearly indicates that the kinetic energy
produces the largest contribution to the dynamics of magnets at finite
temperature while the exchange-correlation contribution is small.

The author would like to thank K. Belashchenko, M. van Schilfgaarde and N.
Zein for helpful discussions. This work was carried out at the Ames
Laboratory, which is operated for the U.S.Department of Energy by Iowa State
University under Contract No. W-7405-82. This work was supported by the
Director for Energy Research, Office of Basic Energy Sciences of the
U.S.Department of Energy.

\clearpage

\section{Figures captions}

Fig. 1. Magnetization of bcc Fe as a function of temperature obtained for
the simulation cell containing 128 atoms with dynamic spin-spiral boundary
conditions. Black stars and white circles correspond to orbital and the
rigid spin approximation spin dynamics, respectively. Black squares show
experimental data. The arrow indicates the finite size scaling result for
the Curie temperature of Fe.

Fig. 2. Total energy as a function of angle between the magnetic moment of a
single isolated impurity and the global bulk magnetization. Black circles
correspond to a Ni impurity in fcc Ni; triangles, to a Fe impurity in bcc
Fe; white circles, to a Ni atom with the relaxation of its nearest neighbors
included. A function $(1-cos\theta )$ is also shown by squares. The energy
is normalized by the stiffness of the corresponding problem. Arrows indicate
the `critical' angles where the local moment on the given impurity
disappears.

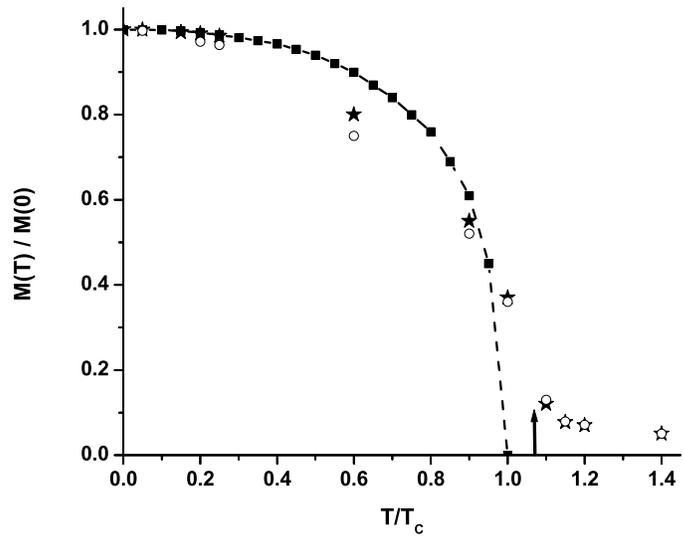
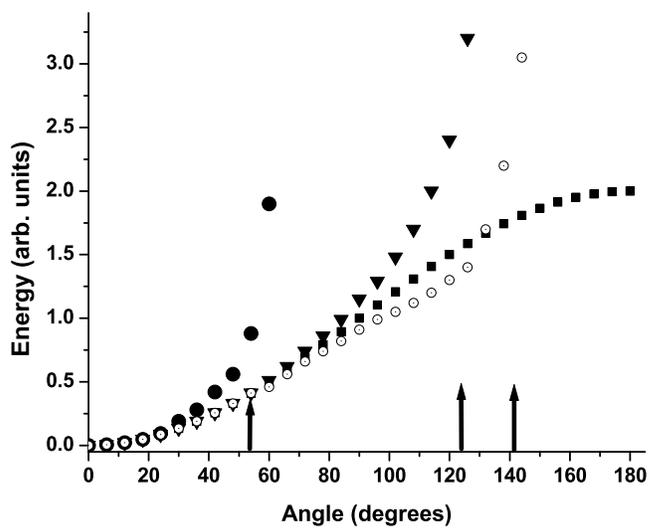
\end{document}